\documentclass{template/WileyChemistry-template}
\usepackage{multirow}
\usepackage{color,soul}

\title{\raggedright Multiple Adsorption of CO Molecules on Transition Metal Substitutional Impurities in Copper Surfaces}

\author{
\begin{minipage}{\textwidth}
	Magnus A. H. Christiansen,*\textsuperscript{+,[a]} Wei Wang,\textsuperscript{+,[b]} Elvar \"{O}. J\'{o}nsson,\textsuperscript{[a]} Giancarlo Cicero,\textsuperscript{[b]} Hannes J\'{o}nsson*\textsuperscript{[c]}
\end{minipage}
}

\newcommand{\affiliation}{
\begin{itemize}

\item[{[a]}] M. A. H. Christiansen,$^\texttt{+}$ E. \"{O}. J\'{o}nsson\\
Science Institute of the University of Iceland, 107 Reykjavík, Iceland\\
E-mail: mah107@hi.is

\item[{[b]}] W. Wang,$^\texttt{+}$ G. Cicero\\
Department of Applied Science and Technology, Politecnico di Torino, Torino 10129, Italy\\

\item[{[c]}]  H. J\'{o}nsson\\
Faculty of Physical Sciences, University of Iceland, 107 Reykjavík, Iceland\\
On leave at Oxford University, Oxford, UK\\
E-mail: hj@hi.is

\item[{[\texttt{+}]}] These authors contributed equally.
\end{itemize}
}

\renewcommand{\abstract}
{
Copper-based catalysts are of particular interest for electrochemical reduction of CO$_2$ (CO2RR) as products beyond CO can form. To improve activity and selectivity, several studies have focused on the addition of other elements as substitutional impurities. Although the adsorption of a single CO molecule has often been used as a descriptor for CO2RR activity, our recent calculations using the RPBE functional showed that multiple CO molecules can bind to first-row transition metal impurities. Here, we extend the study to second-row transition metals and also to a functional that explicitly includes dispersion interaction, BEEF-vdW. The binding energy of the first CO molecule on the impurity atom is found to be significantly larger than on the clean Cu(111) and Cu(100) surfaces, but the differential binding energy generally drops as more CO molecules adsorb. The dispersion interaction is found to make a significant contribution to the binding energy, in particular for the last and weakest bound CO molecule, the one that is most likely to participate in CO2RR. In some cases, four CO admolecules can bind more strongly on the impurity atom than on the clean copper surface. The adsorption of CO causes the position of the impurity atom to shift outwards and in some cases, even escape from the surface layer. The C-O stretch frequencies are calculated in order to identify possible experimental signatures of multiple CO adsorption.
}

\begin{document}

\twocolumn[\vspace{-1.5cm}\maketitle\vspace{-1cm}
	\textit{\dedication}\vspace{0.4cm}]
\small{\begin{shaded}
		\noindent\abstract
	\end{shaded}
}

\begin{figure} [!b]
\begin{minipage}[t]{\columnwidth}{\rule{\columnwidth}{1pt}\footnotesize{\textsf{\affiliation}}}\end{minipage}
\end{figure}



\section*{Introduction}
Copper is a promising catalyst for CO$_2$ electroreduction (CO2RR), as products beyond CO can form such as methane and ethene.\cite{Hori_2002,Hori_2003}
There, the CO intermediate binds strongly enough on the surface to get further reduced while the coverage of hydrogen remains low because of weak binding to H adatoms, and thereby relatively slow hydrogen evolution reaction.
The CO2RR catalytic activity has been characterized by using a two-parameter descriptor involving the binding energy of a CO admolecule on the one hand and the adsorption strength of hydrogen on the other hand, either at high H-coverage\cite{Hussain_2018} or for a single H adatom.\cite{Bagger_2019}
Although the two are somewhat correlated, there is a qualitative difference in that the dispersion interaction makes an important contribution to the binding of a CO molecule while it is unimportant for the binding of an H atom. Too weak binding of CO leads to its release, and thereby no further reduction, as on silver surfaces,\cite{Hoshi_1997,ReFiorentin_2024} while the presence of H adatoms on the surface can lead to a competing reaction pathway resulting in the release of formate without further reduction.\cite{VandenBossche_2021} 
On surfaces where the hydrogen coverage can become high, such as Pt surfaces, only hydrogen is formed under reducing conditions\cite{Hori_1989,Hori_1994} 
because the rate of the hydrogen evolution reaction is significantly higher than the CO2RR rate.\cite{Hussain_2016} 
The two parameter descriptor of a CO molecule and hydrogen adsorption can, in principle, also be used to predict the efficiency of copper surfaces modified by incorporation of other elements, but additional complications may arise, as it may not be sufficient to consider the binding of only a single CO molecule.\cite{christiansen2024}

The activity and selectivity of copper as a CO2RR catalyst is, however, not high enough for the process to be cost-effective. 
Several experimental reports have shown that the 
addition of another element to a copper surface, such as Co\cite{Bernal_2018,Yan_2021}, Ni\cite{Xu_2020} or Pd\cite{Rahaman_2020,Li_2020,Xie_2022}, can lead to significant improvements. 
Even a trace amount of these atoms can substantially influence the reaction rate, selectivity and overpotential required for CO2RR.
These copper-based surfaces with small amounts of substitutional impurities belong to a class of catalysts known as single atom alloys (SAAs).\cite{hannagan2020single,cui2018bridging,yang2013single,kyriakou2012isolated,reocreux2021one,lee2022dilute,Schumann_2024} There, the secondary metal is present in such a low concentration that the atoms are dispersed as substitutional impurities in the surface of the host metal. Typically, the host metal is a less active but more selective material, copper in the present case, while the substituent atoms are of more reactive elements, here transition metals with less occupied d-electron orbitals.

In computational searches for improved CO2RR electrocatalyst materials, the binding energy of an adsorbed CO molecule to the surface is commonly used as a key indicator for the catalytic activity. According to the Sabatier principle,\cite{sabatier1911hydrogenations} a good catalyst should strike a balance between having strong enough binding energy of intermediates to the surface and weak enough binding energy of the final product for facile release from the surface. The binding energy of a CO molecule is taken to be a reliable descriptor for the various intermediates involving bonding to a C atom to the surface, such as in COOH, COH, CHO, and CH$_3$. Consequently, the search for improved CO2RR catalysts has focused on optimizing the binding energy of a single CO molecule on the catalyst surface.\cite{kolb2021structure, calle2017accounting}
However, recent density functional theory (DFT) calculations on CO binding to first-row transition metal (TM) impurities in Cu(111) and Cu(100) surfaces, show that 
up to four CO molecules can bind to the same TM surface atom in some cases.\cite{christiansen2024} The first CO molecule binds so strongly that it is not likely to participate in CO2RR. Instead, it is either the second, third, or possibly the fourth CO molecule that binds weakly enough to be reactive. It is, therefore, essential to identify the CO molecule that binds weakest to the substituent atom, while the binding energy is still comparable to that which other sites on the surface offer.

Here, we present results of calculations on the binding of multiple CO molecules onto substitutional atoms from the second row of TMs in Cu(100) and Cu(111) surfaces, as well as a re-evaluation of the binding to first-row TMs using a density functional that has been shown to be more accurate for CO binding to copper surfaces and includes explicitly the dispersion interaction. 
Although the dense packed (111) facet corresponds to lowest surface energy, the second lowest energy facet, the (100) surface, is  especially interesting as it has been shown to produce more C$_2$ product, namely ethylene, than C$_1$ products, such as methane.\cite{Schouten_2012}
For all the first and second-row TM, except for Zn, Ag, and Cd, the substitutional atom adsorbs more than one CO molecule more strongly than on the pure copper surface, and in some cases up to four CO molecules.
The binding of multiple CO molecules to the substitutional atom significantly affects its position, in some cases displacing it by over 0.8 Å outward from the surface plane and even, for Mn, Fe, and Co, bringing it out of the lattice site.
Various properties, such as magnetic moment, are evaluated and the vibrational frequencies of the adsorbed CO molecules are calculated to assess the possibility of detecting the multiple CO binding experimentally by Raman or infrared (IR) spectroscopy.


\section*{Methods}

The spin-polarized density functional theory (DFT) calculations are carried out using the Bayesian error estimation functional, BEEF-vdW\cite{Wellendorff_2012}.
It includes a nonlocal correlation term representing fluctuating dipoles (see eq. 3 of ref. \citenum{Wellendorff_2012}), which is referred to as dispersion interaction.
It has been shown to give an estimate of CO adsorption energy on various surfaces that is in good agreement with experiments.\cite{campbell2015}
The calculations are carried out for a four-layer slab with a $4\times4$ surface supercell using a plane wave representation of the valence electrons with a cut-off energy of 450 eV, while the projector augmented wave (PAW) approach \cite{Blochl_1994} is used to describe the effect of the inner electrons.
A $5\times5\times1$ Monkhorst-Pack k-point grid is used.
 
The minimum energy configuration of the atoms is obtained using a conjugate gradient optimizer until all atomic forces are smaller than 0.01 eV/Å, and the electronic structure is optimized to self-consistency to a tolerance of $10^{-6}$ eV.
The relaxed copper crystal has a lattice parameter of 3.6551 {\AA}, 1.39\% larger than the experimental value.
The Vienna ab initio simulation package (VASP)\cite{VASP_3,VASP_4} is used to carry out the calculations.\cite{Kohn_1999}


The differential binding energy, \textit{i.e.} the energy required to remove the n$^\text{th}$ CO molecule from the surface and bring it into the vacuum is calculated as
\begin{equation*}
    E_\mathrm{b\text{-}n^{th}\,CO/M@Cu}  = E_\mathrm{CO} + E_{\mathrm{(CO)_{n\text{-}1}/M@Cu}} 
\end{equation*}    
\begin{equation*}
      \ \ \ \ \ \ \ \ \ \ \ \ \ \ \ - E_{\mathrm{(CO)_{n}/M@Cu}},
\end{equation*}
where M is the substitutional atom and M@Cu is the copper surface with a substitutional atom. $E_{\mathrm{CO}}$ is the energy of the isolated CO molecule, while $E_{\mathrm{(CO)_{n}/M@Cu}}$ 
is the energy of the system consisting of the M@Cu surface with $n$ CO molecules bound to the substituent atom. 
The number of unpaired electrons, i.e. magnetization, is evaluated from the output of the spin-polarized DFT calculations, calculated from the difference in integrated spin-up and spin-down electron density.

Only CO adsorption is considered here and the presence of adsorbed hydrogen atoms not considered.
The hydrogen evolution reaction is a competing reaction and the coverage of H adatoms is likely to be significant under CO2RR conditions. The H adatoms compete with CO molecules for surface sites. 
This would add significant complexity to the calculations, as multiple configurations of H and CO on the surface should be evaluated to determine the minimum energy arrangement for different levels of hydrogen coverage and varying numbers of adsorbed CO molecules. This will be addressed in future studies.


The data on which the results presented in this article are based and the input parameters for the calculations are available at Zenodo.\cite{zenodo_repository}

\section*{Results and Discussion}

\subsection*{CO adsorption on Cu(111) and Cu(100)}

The Cu(100) surface has three adsorption sites: top, bridge, and hollow site.  
Calculations with the BEEF-vdW functional give strongest CO binding on the top site, followed by the bridge site, and then the hollow site provides the weakest binding. 
The top site binding energy is estimated to be 0.61 eV based on the minimum energy configuration, and this gets reduced to 0.57 eV if the zero-point energy is subtracted. This is in good agreement with the experimentally determined value of 0.53 eV.\cite{Vollmer_2001, truong_1992} 
The ordering of the binding sites is also in agreement with both the interpretation of the experimental measurements and results of higher-level theoretical calculations based on the random phase approximation.\cite{wei_2021}
The commonly used RPBE functional does not give the same ordering, as it predicts the hollow site to provide the strongest binding to the CO molecule.

The Cu(111) surface has four distinct adsorption sites: top, bridge, hollow FCC, and hollow HCP sites. The BEEF-vdW calculations give strongest CO binding at the hollow FCC site, followed by the hollow HCP site, then the top site, and finally the bridge site. The binding energy at the hollow site is 0.53 eV based on the minimum energy configuration and 0.49 eV if the zero-point energy is subtracted, in close agreement with the experimental estimate of 0.49 eV.\cite{Vollmer_2001}


\subsection*{CO binding on 3d TM impurity}

\begin{figure}
\begin{center}
\includegraphics[width=8.cm]{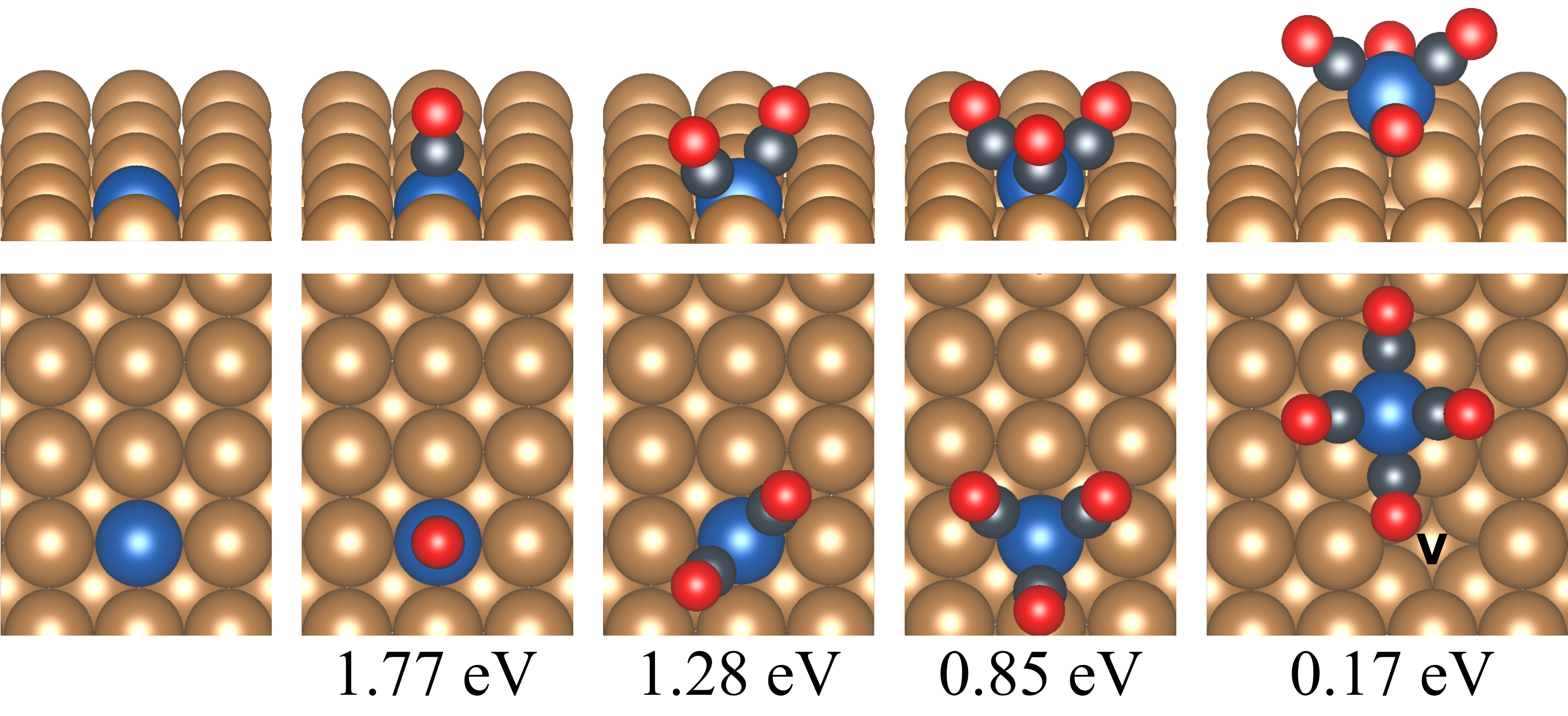}
\caption{
Minimum energy atomic structure of the Cu(100) surface with a Co substitutional atom (blue) and up to four CO molecules 
(black: C atoms; red: O atoms).
After the fourth CO molecule is adsorbed, the Co atom moves out of the surface layer and a vacancy (indicated by V) is formed.
The differential binding energy is written below their respective structures.
This is a composite figure, the calculations are carried out for just one substitutional atom in the surface.
}
\label{fig:CoCu(100)}
\end{center}
\end{figure}

The introduction of a metal atom from the first-row of the transition metals to the left of copper into the Cu(111) and Cu(100) surfaces creates a strong adsorption site for CO molecules. 
An example is shown in Figure \ref{fig:CoCu(100)}, where a Co atom has replaced one of the Cu atoms in the Cu(100) surface.
The minimum energy structures of up to four CO adsorbed to the Co atom are shown. As more CO molecules bind to the Co atom, its position moves outwards from the plane of the surface atoms. When four CO molecules are adsorbed, the Co
atom moves out of the surface site, and a vacancy is formed in the surface layer.
The differential binding energy of the fourth CO molecule is, however, smaller than the binding energy to the clean Cu(100) surfaces, so this displacement of the Co(CO)$_4$ complex will only occur at high CO coverage.


\begin{figure}[h]
\begin{center}
\includegraphics[width=8.cm]{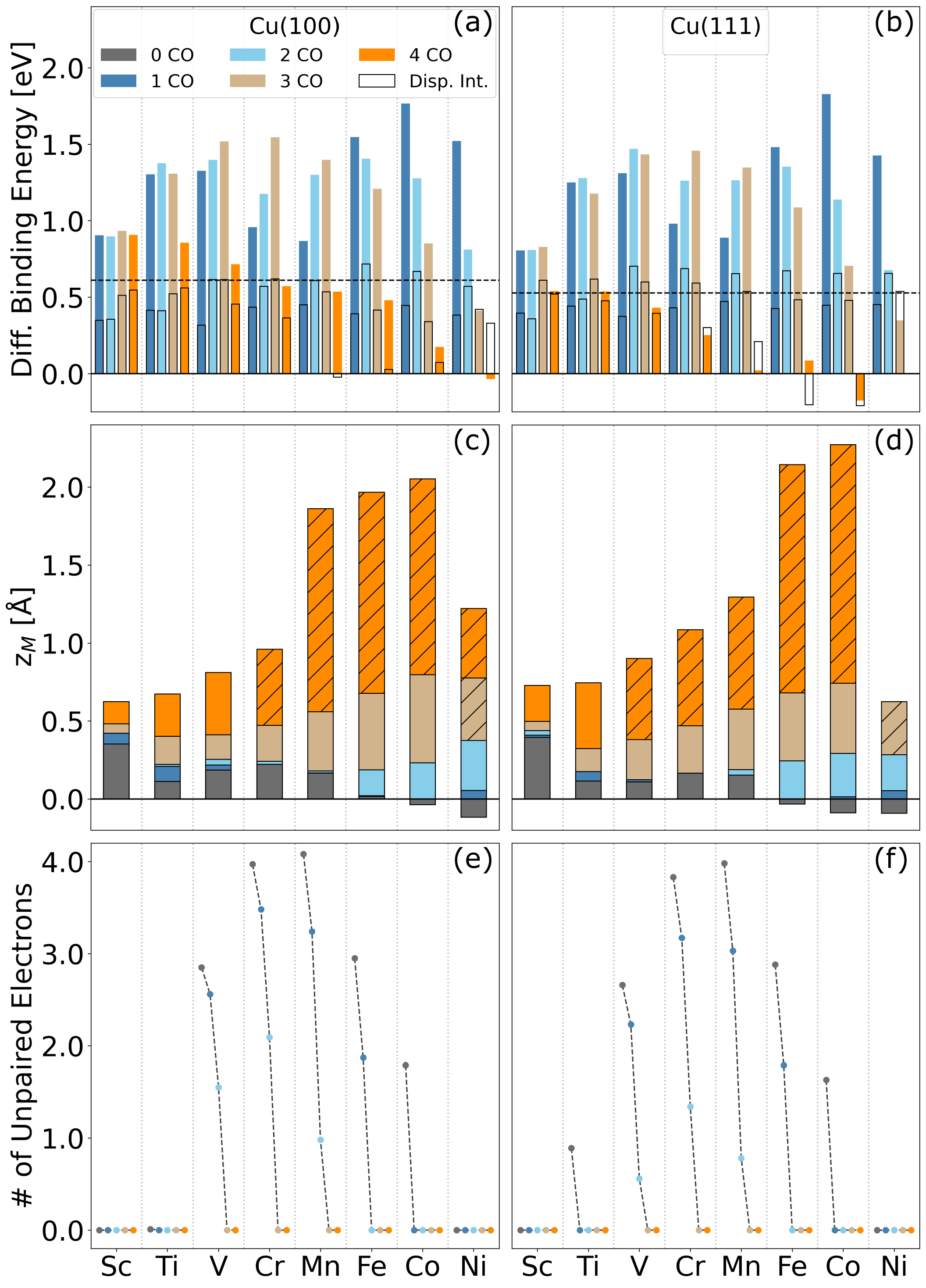}
\caption{Left panel shows calculated results for Cu(100) and the right panel for Cu(111).
(a, b): Differential binding energy, $E_\mathrm{b\text{-}n^{th}\,CO/M@Cu}$, for first, second, third and fourth CO molecule bound to a 3d transition metal substitutional atom.
The open bars indicate the contribution from the dispersion interaction as estimated by the BEEF-vdW functional.
The black dashed lines indicate the binding energy of a single CO molecule on the clean copper surface. 
(c, d): z-Coordinate of the M atom with respect to the copper surface which lies in the xy-plane at $z=0$ when up to four CO molecules adsorb to the M atom. The hatched bars indicate structures that are higher in energy than a structure where one of the CO molecules is placed on the pure copper surface, \textit{i.e.}, far from the M atom.
(e, f): Number of unpaired electrons of the substitutional M atom. The binding of CO increases the energy splitting between the d-orbitals and reduces the number of unpaired electrons.
}
\label{fig:TM1_dE_z}
\end{center}
\end{figure}

The calculated differential CO binding energy
on a 3d TM substitutional atom (Sc, Ti, V, Cr, Mn, Fe, Co, Ni) in Cu(100) and Cu(111) surfaces is illustrated in Figure \ref{fig:TM1_dE_z}.
The binding energy of CO on the clean copper surfaces is indicated for reference.
The results obtained here using the BEEF-vdW functional are similar to the previously reported results obtained with the
RPBE functional,\cite{christiansen2024} 
although it does not include the dispersion interaction.
Since the RPBE functional is fitted to give quite good binding energy for molecules on surfaces even without explicit inclusion of dispersion interaction, it is implicitly taking the net effect of the dispersion interaction into account in some way.

Three distinct trends in the differential binding energy can be seen from Figure \ref{fig:TM1_dE_z}.
The expected trend of decreasing bond strength with the number of CO molecules is seen for
Fe, Co, and Ni substitutional atoms.
As a general rule, an atom makes weaker bonds the more bonds it forms.
For Fe and Co, the differential binding energy is stronger for the first, second, and third adsorbed CO molecules compared to the pure Cu surface on both the (100) and (111) facets. A metastable minimum energy structure is obtained for the fourth adsorbed CO molecule, although the differential binding energy at that point is weaker than for adsorption on the pure copper surface, \textit{i.e.}, far from the substitutional atom. For Ni, the differential binding energy is stronger for the first and second adsorbed CO molecules compared to the pure surface of Cu(100) and Cu(111), while the third and fourth COs binds stronger to the pure copper surfaces.

The opposite trend, 
namely, increasing differential binding energy, is obtained for the first three CO molecules binding to V, Cr, and Mn substitutional atoms. This trend stops for the fourth CO molecule as the differential binding energy drops abruptly. On the Cu(100) surface, the fourth CO binds slightly weaker to the Cr and Mn than to the pure copper surface, by 0.04 and 0.08 eV, respectively. For the Cu(111) surface, the fourth CO binds significantly weaker to the Cr and Mn compared to the pure copper surface, by 0.28 eV and 0.51 eV, respectively.

The third trend is roughly constant differential binding energy for the first three CO molecules on Ti and all four on Sc. When placed in the Cu(100) surface, the differential binding energy of the fourth CO is still significantly stronger for the Sc and Ti compared to the pure copper surface. In the Cu(111) surface, the differential binding energy for the fourth CO is almost identical (ca. 0.01 eV difference) to the CO binding energy on Cu(111).

For the V substitutional atom, the trend is somewhat different depending on which copper facet it is placed in. For the Cu(100) surface, binding to a V substitutional atom shows a similar trend as for Cr and Mn atoms, \textit{i.e.}, increased differential binding energy for the first three adsorbed CO molecules and then a decrease for the fourth CO, but still stronger binding than to the pure copper surface. When placed in the Cu(111) surface, the differential binding energy decreases slightly for the third CO. The fourth CO binds weaker to the V atom than on the clean Cu(111) surface.

Figure \ref{fig:TM1_dE_z} also shows how the position of the substitutional metal atom relative to the copper surface plane changes as the CO molecules are adsorbed. Sc, the largest of the first-row transition metal atoms, initially sits 0.35 and 0.39 Å above the Cu(100) and Cu(111) surface planes, respectively, in the absence of CO adsorption. As the CO molecules adsorb, the Sc atom is further displaced outward by 0.2-0.3  Å.
The smallest of these atoms, Ni, initially sits 
$\sim$0.1 Å below the Cu(100) and Cu(111) surface planes. Co exhibits similar behavior, sitting 0.04 and 0.09 Å below the Cu(100) and Cu(111) surface planes, respectively.
The V atom exhibits the largest 
outward displacement, 0.81 Å above the Cu(100) surface with four adsorbed CO molecules. On the Cu(111) surface, the Ti atom shows the largest displacement, 0.75 Å above the surface plane with four CO molecules adsorbed.

A more drastic displacement occurs when four CO 
are bound to Mn, Fe, and Co substitutional atoms in the Cu(100) surface,
as they move from the lattice site and relocate onto the copper surface, thereby creating a vacancy in the surface layer. 
The Mn, Fe, and Co atoms are positioned 1.86, 1.97, and 2.05 Å above the Cu(100) surface plane, respectively. 
On the Cu(111) surface, only the Fe and Co substitutional atoms move outward onto the surface, positioned at 2.15 Å and 2.27 Å above the Cu(111) surface plane, respectively.
However, the adsorption of the fourth CO molecule on these atoms is less energetically favorable compared to CO adsorption on the pure Cu(100) surface, as shown in Figure \ref{fig:TM1_dE_z}a. Therefore, these structures are only likely to form for high CO coverage.

The open bars in Figure \ref{fig:TM1_dE_z}a,b indicate the contribution from the dispersion interaction to the differential binding energy as obtained from the BEEF-vdW functional. As seen with the Sc substituent atom, which undergoes only small change in its position with respect to the surface plane as CO molecules adsorb, the contribution of the dispersion interaction to the differential binding energy tends to increase with the number of adsorbed CO molecules. By the time the fourth CO molecule adsorbs, the majority of the binding energy arises from dispersion interaction. 
An exception to this trend is observed for the Mn, Fe, and Co substituent atoms, where the interaction between the CO molecules and the substituent atoms is so strong that the atoms are pulled from their lattice sites onto the copper surface. As a result, the distance between the copper surface and the metal-carbonyl complex increases significantly, leading to a decrease in the dispersion interaction. In these cases, the differential dispersion interaction contribution even becomes negative.

The change in the differential binding energy of the CO molecules can be correlated with the number of unpaired d-electrons.
Figure \ref{fig:TM1_dE_z}e,f shows the calculated magnetic moments of the substitutional transition metals atoms. For Sc and Ni, the initial magnetic moment is zero, while Ti and Co begin with a small magnetic moment that disappears upon adsorption of the first CO molecule. For the other first-row transition metals, the magnetic moment decreases monotonically in a stepwise manner as more CO molecules adsorb. This results from an increased energy difference between the different d-orbitals of the substitutional atoms as more of the CO molecules adsorb, and thereby more pairing of the d-electrons. This helps explain the unusual increase in differential binding energy calculated for Cr and Mn as the second and third CO molecules are added.
\newline
\newline


\subsection*{CO binding to 4d TM impurities}

\begin{figure}
\begin{center}
\includegraphics[width=8.6cm]{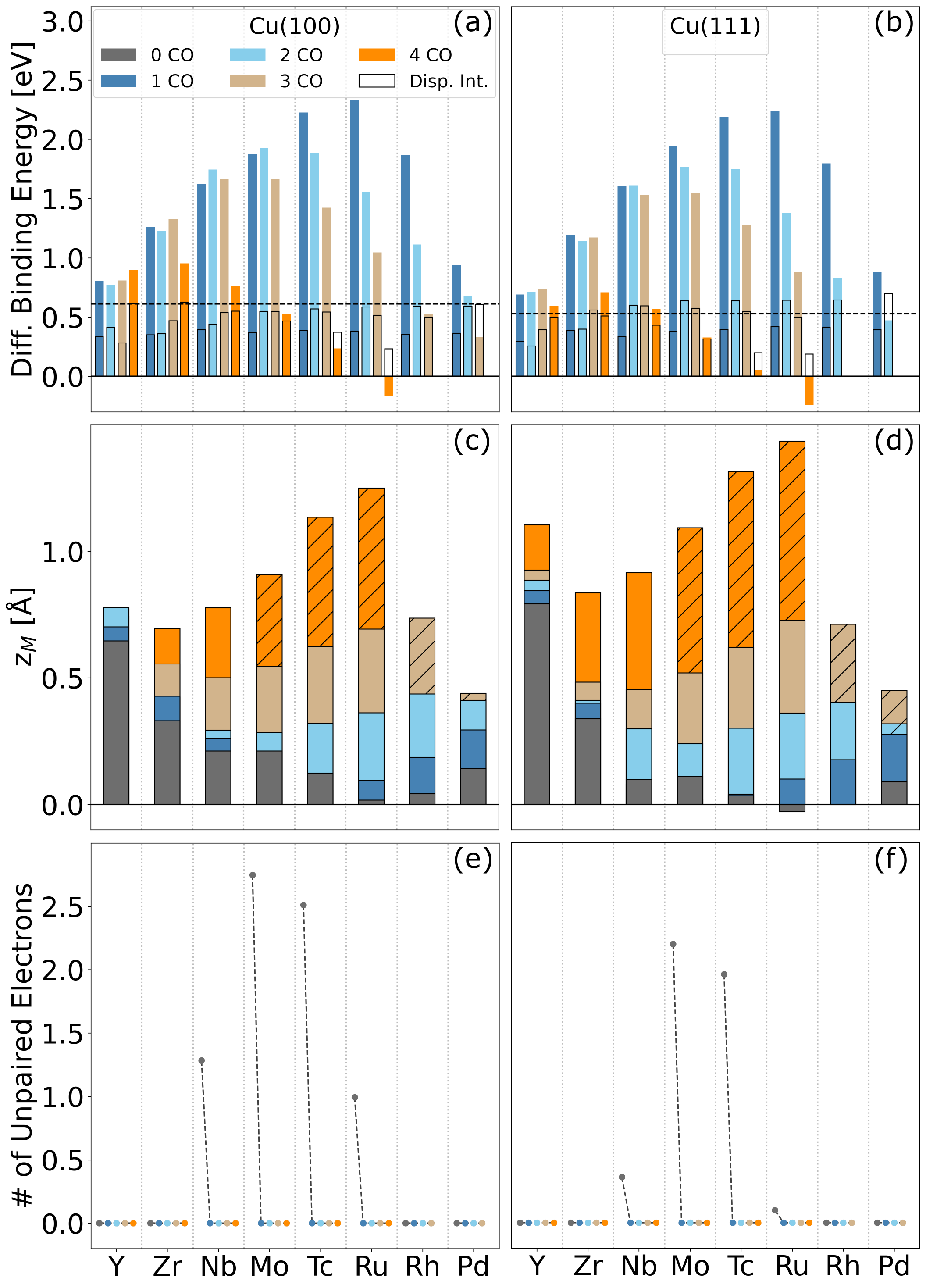}
\caption{
Left panel shows calculated results for Cu(100) and the right panel for Cu(111).
(a, b): Differential binding energy, $E_\mathrm{b\text{-}n^{th}\,CO/M@Cu}$, for first and up to fourth CO molecule bound to a 4d transition metal substitutional atom.
The black dashed lines indicate the binding energy of a single CO molecule on the clean copper surface. The open bar (black outlines) indicates the contribution of the dispersion interaction to the differential binding energy.
(c, d): z-Coordinate of the substitutional atom with respect to the copper surface plane when up to four CO molecules bind to it. The hatched bars indicate structures that are higher in energy than a structure where one of the CO molecules is placed on the pure copper surface, \textit{i.e.}, far from the M atom.
(e, f): Number of unpaired electrons of the substitutional M atom. The binding of CO increases the energy splitting between the d-orbitals and reduces the number of unpaired electrons.
}
\label{fig:TM2_dE_z}
\end{center}
\end{figure}

Figure \ref{fig:TM2_dE_z} shows the calculated results for substitutional atoms from the second-row of transition metals
(Y, Zr, Nb, Mo, Tc, Ru, Rh, Pd).
Two trends are observed in the differential binding energy. 
For Tc, Ru, Rh, and Pd, it decreases as more CO molecules adsorb. In contrast, for Y, Zr, and Nb, the differential binding energy remains relatively constant until it decreases upon adsorption of the fourth CO molecule. A trend of increasing differential binding energy, as obtained for some of the 3d TM substitutional atoms, is not found here.
The first CO molecule is bound most strongly to a Ru atom, the second CO molecule binds most strongly to the Mo atom, the third CO molecule adsorbs most strongly to the Nb and Mo substituent atoms, while the fourth CO molecule is most strongly bound to the Zr substitutional atom.

The Mo, Tc, and Ru atoms can adsorb up to three CO molecules with a differential binding energy stronger than that of CO adsorption on pure copper surfaces, while Rh adsorbs up to two CO molecules. Pd adsorbs two CO molecules when placed in the Cu(100) surface and one when placed in the Cu(111) surface more strongly than the clean copper surface. Y, Zr, and Nb can adsorb up to four CO molecules with a differential binding energy that exceeds that of CO adsorption on the pure copper surfaces.
Among the 4d TMs on the Cu(100) surface, Mo, Rh, and Pd show the most promise for enhanced CO2RR. The differential binding energy for the fourth CO bound to Mo and for the third CO bound to Rh in the Cu(100) surface are both lower by ca. 0.08 eV, compared to the CO binding energy on clean Cu(100). The second CO binds to Pd atom in the Cu(100) surface with a differential binding energy  0.07 eV stronger than the binding energy of CO on the Cu(100) surface, so it is also likely to be able to participate in CO2RR.

Figure \ref{fig:TM2_dE_z}c,d shows the position of the 4d TM substitutional atoms with respect to the plane of the copper surface atoms.
The atom with the largest radius, Y, sits above the surface plane by 0.6 and 0.8 Å for Cu(100) and Cu(111), respectively. As the atomic radius gets reduced going to the right along the second row of TM, the substitutional atom sits closer to the surface. The Ru atom is nearly in the surface plane of Cu(100) while it is slightly below the surface when placed in the Cu(111) surface.
Compared to their first-row counterparts, significant differences in the positions are found only for Y, Zr, and Pd.
For Nb, Mo, Tc, Ru, and Rh, the difference compared to their first-row counterparts is less than 0.1 {\AA}.

The binding of CO molecules to the Y substitutional atom does not change much its location over the surface plane.
When placed in the Cu(100) surface, Y is displaced by up to 0.13 Å, with no significant changes after the first two CO molecules adsorb. On the Cu(111) surface, Y is displaced outwards by 0.31 Å with four adsorbed CO molecules.
For Mo, Tc, and Ru, which are smaller than Y, the substitutional atom is displaced slightly in the absence of CO, while the adsorption of the fourth CO molecule can cause a significant outward displacement of over 1 Å. 
On both the Cu(100) and Cu(111) surfaces, Nb, and Ru exhibit the largest displacements as they bind to CO molecules. 
With three CO molecules adsorbed, Ru is displaced by an additional 0.67 Å on the Cu(100) surface and 0.76 Å on the Cu(111) surface. In the case of Nb bound to four CO molecules, the outward displacement is 0.57 Å when placed in the Cu(100) and by 0.82 Å when placed in the Cu(111) surface. 

None of the 4d TM substitutional atoms are found to move onto the copper surface to form an adsorbed carbonyl complex and create a vacancy in the surface layer as found for the Mn, Fe, and Co substitutional atoms when a fourth CO is adsorbed. This indicates that 4d TM substitutional atoms might be more stable as SAA catalysts compared to their 3d TM counterparts.

The number of unpaired electrons of the 4d TM substitutional atoms is smaller 
than for corresponding 3d TMs, as shown in Figure \ref{fig:TM2_dE_z}e,f. For Nb, Mo, Tc, and Ru, the magnetic moment drops to zero after adsorbing just a single CO molecule, in contrast to the stepwise decrease observed for the 3d TMs. As can be seen from the larger differential binding energy of the first CO bound to a 4d TM substitutional atom as compared to the corresponding 3d TM atom, the TM-CO interaction is stronger and this leads to a large energy difference between the different d orbitals and thereby an increased tendency for pairing of the electrons.

\begin{table*}
	\begin{center}
	\caption{
    CO stretching frequencies ($\text{cm}^{-1}$) for CO molecules bound to a substitutional atom, M, in the Cu(100) surface, as well as an isolated CO molecule and CO adsorbed on the pure Cu(100) surface.
    }\label{tab1}
		\begin{tabular}{lccc}	
\toprule
M@Cu(100) & Configuration & Frequency ($\text{cm}^{-1}$) & Reference \\
\midrule
 -	& CO (gas) & 2127 & 2143 (exp)\textsuperscript{[a]}\\ \cline{1-4}
 Cu & CO(T) & 2008 & 2082-2112 (exp), 2011 (calc)\textsuperscript{[b]}\\
    & CO(B) & 1877 & 1872 (calc)\textsuperscript{[c]} \\
    & CO(H) & 1696 &\\ \cline{1-4}
Rh 	& 1 CO	 & 1955 & 1978 (exp), 1934 (calc)\textsuperscript{[d]}\\
    & 2 CO   & 1896, 1864	& 1926 (exp), 1891 (calc) \textsuperscript{[d]}\\
    & 3 CO	 & 1897, 1862, 1860   &	\\ \cline{1-4}
Pd 	& 1 CO   & 2004 & 2032 (exp, low exposure)\textsuperscript{[e]}\\
    & 2 CO   & 1910, 1882 & \multirow[c]{2}{*}{1887 (exp, high exposure)\textsuperscript{[e]}}\\
    & 3 CO	 & 1928, 1899, 1728 &	\\ \cline{1-4}
V 	& 1 CO   & 1964 &	\\
    & 2 CO   & 1861, 1831 &	\\
    & 3 CO	 & 1852, 1821, 1788 &	\\
    & 4 CO	 & 1862, 1816, 1816, 1812 & \\ \cline{1-4} 
Co 	& 1 CO   & 1929 & \\
    & 2 CO   & 1846, 1819  & \\
    & 3 CO   & 1891, 1852, 1847 &	\\
    & 4 CO	 &2017, 1982, 1853, 1824 &	\\
\bottomrule	
	\end{tabular}
	\end{center}
\footnotesize{\textsf{
[a] Experimental result from Huber et al.\cite{huber_1979} 
[b] Experimental and calculated results from Roiaz et al.\cite{Roiaz_2019} 
[c] Calculated result from Zhan et al.\cite{zhan_2024} 
[d] Experimental and calculated results from Wang et al.\cite{Wang_2022} 
[e] Experimental results on CuPd alloy with 9\% Pd from Rao et al.\cite{Rao_1985}.}}

\end{table*}

\subsection*{CO stretch vibrational frequency}

The C-O vibrational frequency has been calculated within the harmonic approximation for some of the systems in order to identify possible experimental signatures of multiple CO adsorption on substitutional TM atoms. 
The results are shown in Table \ref{tab1} along with calculated results for gas phase CO and CO adsorbed on Cu(100).
Some experimental data and previous vibrational calculations are available for some of the systems and are listed in the table for
comparison.
Although the calculated frequency differs from the experimental results for gas phase CO and CO adsorbed on Cu(100), possibly because of anharmonic contributions as well as shortcomings of the functional approximation, 
the trends for different binding sites agree well with experiments.\cite{Roiaz_2019} 

The CO stretch frequency decreases upon adsorption on the Cu(100) surface due to back-donation from the metal to the $\pi^*$ orbital  of the CO molecule, thereby weakening the C–O bond and lowering its stretch frequency.\cite{Goldman_1996,Bistoni_2016}
The stronger this back-donation interaction is, the lower the CO stretching frequency becomes.  
 A Rh substitutional atom in Cu(100) surface has been shown to adsorb two CO molecules, both from experiments and DFT calculations. The monocarbonyl and symmetric dicarbonyl stretches appear at 1978 and 1926 $\text{cm}^{-1}$ with a 52 $\text{cm}^{-1}$ interval according to experiments, consistent with the calculated results as shown in Table \ref{tab1}.\cite{Wang_2022} Our results give 1955 and 1896 $\text{cm}^{-1}$ with a difference of 59 $\text{cm}^{-1}$, in good agreement with the experimental results. 
 
 For a CuPd alloy, experimental measurements\cite{Rao_1985} found CO stretching frequencies at 2032 $\text{cm}^{-1}$ at low CO exposure and 1887 $\text{cm}^{-1}$ at high CO exposure, which is close to our calculations of 2004 $\text{cm}^{-1}$ for one CO adsorption and \textit{ca.} 1920 $\text{cm}^{-1}$ for two and three CO adsorptions, indicating that the Pd atoms on the alloy surface were bound to two CO molecules in the experimental measurements. The difference in calculated stretch frequencies for two and three CO molecules adsorbed on substitutional atoms is small, so it may be hard to distinguish these two configurations by vibration spectroscopy. 
 
 For a V substitutional atom, which can adsorb up to four CO molecules more strongly than the pure Cu(100) surface, the difference between monocarbonyl and dicarbonyl stretches is over 100 $\text{cm}^{-1}$, and should be observable experimentally. The symmetric CO stretches for multiple CO adsorption are, however, close, and there is only a slight red shift from two to three CO molecules and a blue shift from three to four, consistent with the changes in calculated differential binding energy. 
 The asymmetric CO stretch of three CO molecules on the V atom at 1788 $\text{cm}^{-1}$ may offer a measurable signature. 
 
 On a Co substitutional atom, the calculated results show a significant red shift from one to two adsorbed CO molecules, followed by a slight blue shift with further CO adsorption. This is a different trend from the assignments made in experimental studies\cite{Eren_2018} of Co deposited on the Cu(110) surface, possibly due to clustering of the Co atoms and/or the broad infrared absorption band.   


\section*{Conclusions}
The results presented here from calculations of binding of up to four CO molecule on 3d and 4d TM substitutional atoms in Cu(100) and Cu(111) surfaces using the BEEF-vdW functional highlight the importance of multiple CO adsorption and the importance of the dispersion interaction, especially for the last and weakest bound CO molecule, which is expected to be most important for CO2RR. 
The differential binding energy decreases as more CO molecules adsorb for most 3d and 4d TM substituent atoms, but binding to Cr and Mn atoms shows the opposite trend which can be attributed to a change in the number of unpaired d-electrons due to the ligand effect.
When Mn, Fe, and Co substitutional atoms bind four CO molecules, they leave the lattice site and form a carbonyl complex on the copper surface indicating a possible degradation mechanism of SAA catalysts.
Calculated CO stretch frequencies show that the adsorption of multiple CO molecules could be experimentally detectable with vibrational spectroscopy.

The CO molecules that bind to a substitutional atom with similar strength as CO adsorption to the pure copper surface are most likely to be active reactants in CO2RR.
For the Cu(100) surface, the V, Cr, Mn, Nb, and Mo substitutional atoms bind the fourth CO molecule with a differential binding energy close to that of the pure copper surface, while the third CO binds with similar strength to a Rh atom, and the second CO on a Pd atom.
For the Cu(111) surface, the Sc, Ti, Y, Zr, and Nb substitutional atoms bind the fourth CO with a differential binding energy close to that of the pristine Cu(111) surface. It is unlikely that two CO molecules adsorbed to the same substituent will dimerize, since this would require overcoming the binding energy of both the last and the second-to-last CO (which can be greater than 1.5 eV). However, if the binding energy of the last adsorbed CO is weak enough, it could undergo C–C coupling with a second CO adsorbed on a neighboring site.

Although we have focused here entirely on the reduction of CO, it could also be interesting to study how substitutional atoms in copper could enhance the CO$_2$-to-CO conversion. Also, here we have focused entirely on a single substitutional atom in a copper surface, while there are some indications that some substitutional atoms tend to agglomerate into dimers or larger clusters, depending on concentration.\cite{Eren_2018, han2021single}

Our results show that multiple adsorption of CO molecules on substitutional TM atoms can play a crucial role in determining the activity and stability of copper based electrocatalysts. More generally, similar multiple reactant adsorption could play a role on various SAA catalysts.


\section*{Acknowledgements}
This work was funded by the Icelandic Research Fund (grant no. 207283-053) and the EU’s Horizon 2021 programme under the Marie Skłodowska-Curie Doctoral Networks (MSCA-DN) grant agreement No 101072830 (ECOMATES). H.J. acknowledges support from the Leverhulme Trust (no. VP1-2024-007) and helpful discussions with Prof. Michail Stamatakis.

Views and opinions expressed are those of the authors and do not necessarily reflect the official policy or position of the European Union or the Granting Authority, and neither can be held responsible for them.

\section*{Conflict of Interest}
The authors declare no conflict of interest.

\section*{Data Availability Statement}
The data that support the findings of this study are available in the supplementary material of this article and in Ref. [35].

\begin{shaded}
\noindent\textsf{\textbf{Keywords: CORR, single-atom-alloys, substitutional impurities, Cu(100), Cu(111), Geminal multicarbonyl}} 
\end{shaded}


\setlength{\bibsep}{0.0cm}
\bibliographystyle{template/Wiley-chemistry}
\bibliography{references}

\clearpage


\section*{Entry for the Table of Contents}


\noindent\rule{11cm}{2pt}
\begin{minipage}{11cm}
\includegraphics[width=11cm]{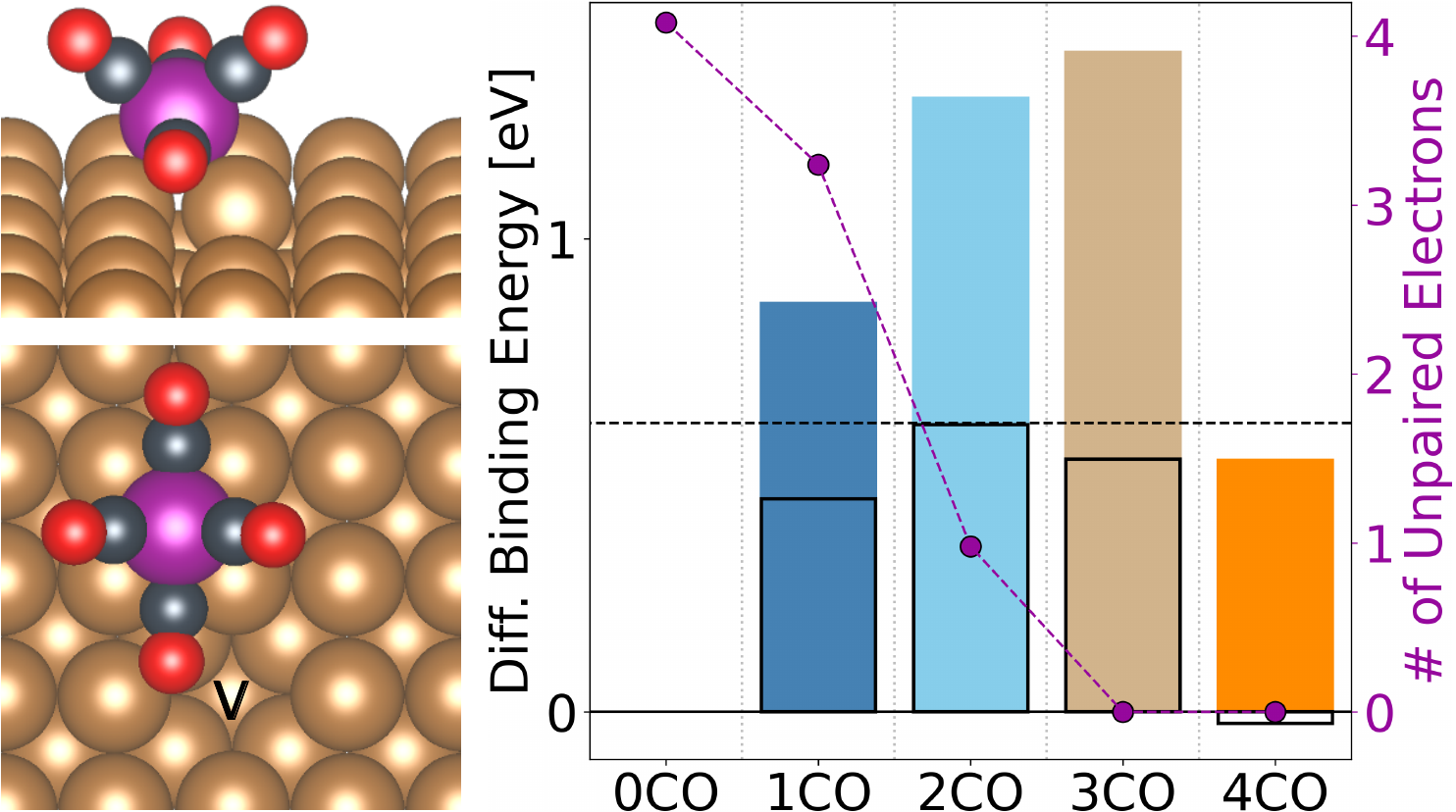}
\end{minipage}
\begin{minipage}{11cm}
\large\textsf{
\\
Adsorption of CO molecules on an Mn impurity atom in the Cu(100) surface. The differential binding energy, including contribution from dispersion interaction (open bars), increases for the first three CO (dashed line indicates CO binding energy for clean surface), as the number of unpaired electrons drops. Formation of Mn(CO)$_4$ results in a displacement from the surface site and formation of a vacancy (V). 
}
\end{minipage}
\noindent\rule{11cm}{2pt}


\end{document}



\vspace{-1.5cm}
\maketitle\vspace{-1cm}

\begin{figure} [!b]
\begin{minipage}[t]{\columnwidth}{\rule{\columnwidth}{1pt}\footnotesize{\textsf{\affiliation}}}\end{minipage}
\end{figure}



\begin{table}[]
\centering
\begin{tabular}{|c|c|c|c|c|c|c|} \hline
M & \#CO & $E_\mathrm{b\text{-}n^{th}\,CO/M@Cu(100)}$ & $E_\mathrm{vdW}$ & $z_M$ & magmom \\ \hline
\multirow{5}{*}{Sc}
& 0 & -- & --  & 0.353 \AA & 0.000 \\
& 1 & 0.905 eV & 0.349 eV & 0.421 \AA & 0.000 \\
& 2 & 0.897 eV & 0.355 eV & 0.418 \AA & 0.000 \\
& 3 & 0.933 eV & 0.512 eV & 0.482 \AA & 0.000 \\
& 4 & 0.908 eV & 0.546 eV & 0.624 \AA & 0.000 \\ \hline
\multirow{5}{*}{Ti}
& 0 & -- & --  & 0.112 \AA & 0.010 \\
& 1 & 1.304 eV & 0.414 eV & 0.209 \AA & 0.000 \\
& 2 & 1.377 eV & 0.412 eV & 0.223 \AA & 0.000 \\
& 3 & 1.307 eV & 0.522 eV & 0.402 \AA & 0.000 \\
& 4 & 0.856 eV & 0.561 eV & 0.673 \AA & 0.000 \\ \hline
\multirow{5}{*}{V}
& 0 & -- & --  & 0.186 \AA & 2.850 \\
& 1 & 1.326 eV & 0.317 eV & 0.218 \AA & 2.560 \\
& 2 & 1.397 eV & 0.615 eV & 0.254 \AA & 1.550 \\
& 3 & 1.518 eV & 0.614 eV & 0.412 \AA & 0.000 \\
& 4 & 0.716 eV & 0.454 eV & 0.812 \AA & 0.000 \\ \hline
\multirow{5}{*}{Cr}
& 0 & -- & --  & 0.222 \AA & 3.970 \\
& 1 & 0.957 eV & 0.434 eV & 0.221 \AA & 3.480 \\
& 2 & 1.175 eV & 0.570 eV & 0.242 \AA & 2.090 \\
& 3 & 1.545 eV & 0.619 eV & 0.473 \AA & 0.000 \\
& 4 & 0.572 eV & 0.364 eV & 0.960 \AA & 0.000 \\ \hline
\multirow{5}{*}{Mn}
& 0 & -- & --  & 0.166 \AA & 4.080 \\
& 1 & 0.867 eV & 0.450 eV & 0.168 \AA & 3.240 \\
& 2 & 1.300 eV & 0.607 eV & 0.180 \AA & 0.980 \\
& 3 & 1.397 eV & 0.534 eV & 0.560 \AA & 0.000 \\
& 4 & 0.535 eV & -0.024 eV & 1.861 \AA & 0.000 \\ \hline
\multirow{5}{*}{Fe}
& 0 & -- & --  & 0.017 \AA & 2.950 \\
& 1 & 1.547 eV & 0.391 eV & 0.021 \AA & 1.870 \\
& 2 & 1.405 eV & 0.717 eV & 0.188 \AA & 0.000 \\
& 3 & 1.209 eV & 0.416 eV & 0.677 \AA & 0.000 \\
& 4 & 0.480 eV & 0.026 eV & 1.966 \AA & 0.000 \\ \hline
\multirow{5}{*}{Co}
& 0 & -- & --  & -0.037 \AA & 1.790 \\
& 1 & 1.767 eV & 0.445 eV & -0.020 \AA & 0.000 \\
& 2 & 1.277 eV & 0.668 eV & 0.232 \AA & 0.000 \\
& 3 & 0.851 eV & 0.339 eV & 0.798 \AA & 0.000 \\
& 4 & 0.174 eV & 0.072 eV & 2.052 \AA & 0.000 \\ \hline
\multirow{5}{*}{Ni}
& 0 & -- & --  & -0.117 \AA & 0.000 \\
& 1 & 1.521 eV & 0.382 eV & 0.055 \AA & 0.000 \\
& 2 & 0.812 eV & 0.570 eV & 0.376 \AA & 0.000 \\
& 3 & 0.411 eV & 0.419 eV & 0.776 \AA & 0.000 \\
& 4 & -0.036 eV & 0.329 eV & 1.221 \AA & 0.000 \\ \hline
\end{tabular}
\caption{Differential binding energy, vdW correction contribution, outward displacement, and magnetic moment (in unpaired electron units) for 3d transition metal substituent atoms (M) on the Cu(100) surface with up to 4 CO molecules adsorbed.}
\label{tab:Cu(100)_3d}
\end{table}

\begin{table}[]
\centering
\begin{tabular}{|c|c|c|c|c|c|c|} \hline
M & \#CO & $E_\mathrm{b\text{-}n^{th}\,CO/M@Cu(111)}$ & $E_\mathrm{vdW}$ & $z_M$ & magmom \\ \hline
\multirow{5}{*}{Sc}
& 0 & -- & --  & 0.395 \AA & 0.000 \\
& 1 & 0.806 eV & 0.396 eV & 0.409 \AA & 0.000 \\
& 2 & 0.808 eV & 0.357 eV & 0.438 \AA & 0.000 \\
& 3 & 0.828 eV & 0.611 eV & 0.498 \AA & 0.000 \\
& 4 & 0.541 eV & 0.520 eV & 0.729 \AA & 0.000 \\ \hline
\multirow{5}{*}{Ti}
& 0 & -- & --  & 0.115 \AA & 0.890 \\
& 1 & 1.250 eV & 0.441 eV & 0.176 \AA & 0.000 \\
& 2 & 1.279 eV & 0.487 eV & 0.163 \AA & 0.000 \\
& 3 & 1.176 eV & 0.617 eV & 0.323 \AA & 0.000 \\
& 4 & 0.537 eV & 0.476 eV & 0.745 \AA & 0.000 \\ \hline
\multirow{5}{*}{V}
& 0 & -- & --  & 0.110 \AA & 2.660 \\
& 1 & 1.310 eV & 0.374 eV & 0.124 \AA & 2.230 \\
& 2 & 1.469 eV & 0.701 eV & 0.122 \AA & 0.560 \\
& 3 & 1.434 eV & 0.598 eV & 0.381 \AA & 0.000 \\
& 4 & 0.431 eV & 0.394 eV & 0.901 \AA & 0.000 \\ \hline
\multirow{5}{*}{Cr}
& 0 & -- & --  & 0.166 \AA & 3.830 \\
& 1 & 0.980 eV & 0.429 eV & 0.157 \AA & 3.170 \\
& 2 & 1.261 eV & 0.686 eV & 0.159 \AA & 1.340 \\
& 3 & 1.458 eV & 0.592 eV & 0.470 \AA & 0.000 \\
& 4 & 0.252 eV & 0.301 eV & 1.086 \AA & 0.000 \\ \hline
\multirow{5}{*}{Mn}
& 0 & -- & --  & 0.153 \AA & 3.980 \\
& 1 & 0.889 eV & 0.471 eV & 0.104 \AA & 3.030 \\
& 2 & 1.264 eV & 0.653 eV & 0.189 \AA & 0.780 \\
& 3 & 1.347 eV & 0.537 eV & 0.576 \AA & 0.000 \\
& 4 & 0.021 eV & 0.209 eV & 1.296 \AA & 0.000 \\ \hline
\multirow{5}{*}{Fe}
& 0 & -- & --  & -0.033 \AA & 2.880 \\
& 1 & 1.481 eV & 0.426 eV & -0.006 \AA & 1.790 \\
& 2 & 1.353 eV & 0.672 eV & 0.245 \AA & 0.000 \\
& 3 & 1.087 eV & 0.483 eV & 0.681 \AA & 0.000 \\
& 4 & 0.086 eV & -0.205 eV & 2.145 \AA & 0.000 \\ \hline
\multirow{5}{*}{Co}
& 0 & -- & --  & -0.089 \AA & 1.630 \\
& 1 & 1.827 eV & 0.447 eV & 0.014 \AA & 0.000 \\
& 2 & 1.138 eV & 0.655 eV & 0.292 \AA & 0.000 \\
& 3 & 0.705 eV & 0.478 eV & 0.743 \AA & 0.000 \\
& 4 & -0.177 eV & -0.210 eV & 2.274 \AA & 0.000 \\ \hline
\multirow{4}{*}{Ni}
& 0 & -- & --  & -0.091 \AA & 0.000 \\
& 1 & 1.427 eV & 0.451 eV & 0.053 \AA & 0.000 \\
& 2 & 0.676 eV & 0.655 eV & 0.284 \AA & 0.000 \\
& 3 & 0.348 eV & 0.537 eV & 0.625 \AA & 0.000 \\ \hline
\end{tabular}
\caption{Differential binding energy, vdW correction contribution, outward displacement, and magnetic moment (in unpaired electron units) for 3d transition metal substituent atoms (M) on the Cu(111) surface with up to 4 CO molecules adsorbed.}
\label{tab:Cu(111)_3d}
\end{table}

\begin{table}[]
\centering
\begin{tabular}{|c|c|c|c|c|c|c|} \hline
M & \#CO & $E_\mathrm{b\text{-}n^{th}\,CO/M@Cu(100)}$ & $E_\mathrm{vdW}$ & $z_M$ & magmom \\ \hline
\multirow{5}{*}{Y}
& 0 & -- & --  & 0.645 \AA & 0.000 \\
& 1 & 0.804 eV & 0.335 eV & 0.701 \AA & 0.000 \\
& 2 & 0.766 eV & 0.410 eV & 0.777 \AA & 0.000 \\
& 3 & 0.808 eV & 0.281 eV & 0.743 \AA & 0.000 \\
& 4 & 0.900 eV & 0.612 eV & 0.771 \AA & 0.000 \\ \hline
\multirow{5}{*}{Zr}
& 0 & -- & --  & 0.330 \AA & 0.000 \\
& 1 & 1.262 eV & 0.349 eV & 0.427 \AA & 0.000 \\
& 2 & 1.229 eV & 0.360 eV & 0.423 \AA & 0.000 \\
& 3 & 1.328 eV & 0.468 eV & 0.554 \AA & 0.000 \\
& 4 & 0.954 eV & 0.625 eV & 0.695 \AA & 0.000 \\ \hline
\multirow{5}{*}{Nb}
& 0 & -- & --  & 0.211 \AA & 1.284 \\
& 1 & 1.625 eV & 0.392 eV & 0.261 \AA & 0.000 \\
& 2 & 1.745 eV & 0.439 eV & 0.293 \AA & 0.000 \\
& 3 & 1.661 eV & 0.537 eV & 0.500 \AA & 0.000 \\
& 4 & 0.763 eV & 0.549 eV & 0.776 \AA & 0.000 \\ \hline
\multirow{5}{*}{Mo}
& 0 & -- & --  & 0.211 \AA & 2.749 \\
& 1 & 1.872 eV & 0.370 eV & 0.054 \AA & 0.000 \\
& 2 & 1.925 eV & 0.547 eV & 0.283 \AA & 0.000 \\
& 3 & 1.663 eV & 0.548 eV & 0.545 \AA & 0.000 \\
& 4 & 0.529 eV & 0.465 eV & 0.908 \AA & 0.000 \\ \hline
\multirow{5}{*}{Tc}
& 0 & -- & --  & 0.123 \AA & 2.512 \\
& 1 & 2.227 eV & 0.386 eV & 0.054 \AA & 0.000 \\
& 2 & 1.885 eV & 0.568 eV & 0.319 \AA & 0.000 \\
& 3 & 1.423 eV & 0.543 eV & 0.623 \AA & 0.000 \\
& 4 & 0.234 eV & 0.371 eV & 1.134 \AA & 0.000 \\ \hline
\multirow{5}{*}{Ru}
& 0 & -- & --  & 0.017 \AA & 0.994 \\
& 1 & 2.334 eV & 0.382 eV & 0.094 \AA & 0.000 \\
& 2 & 1.555 eV & 0.584 eV & 0.361 \AA & 0.000 \\
& 3 & 1.045 eV & 0.514 eV & 0.692 \AA & 0.000 \\
& 4 & -0.167 eV & 0.232 eV & 1.249 \AA & 0.000 \\ \hline
\multirow{4}{*}{Rh}
& 0 & -- & --  & 0.042 \AA & 0.000 \\
& 1 & 1.869 eV & 0.351 eV & 0.186 \AA & 0.000 \\
& 2 & 1.111 eV & 0.592 eV & 0.436 \AA & 0.000 \\
& 3 & 0.522 eV & 0.498 eV & 0.735 \AA & 0.000 \\ \hline
\multirow{4}{*}{Pd}
& 0 & -- & --  & 0.141 \AA & 0.000 \\
& 1 & 0.940 eV & 0.362 eV & 0.294 \AA & 0.000 \\
& 2 & 0.682 eV & 0.593 eV & 0.411 \AA & 0.000 \\
& 3 & 0.331 eV & 0.606 eV & 0.438 \AA & 0.000 \\ \hline
\end{tabular}
\caption{Differential binding energy, vdW correction contribution, outward displacement, and magnetic moment (in unpaired electron units) for 4d transition metal substituent atoms (M) on the Cu(100) surface with up to 4 CO molecules adsorbed.}
\label{tab:Cu(100)_4d}
\end{table}

\begin{table}[]
\centering
\begin{tabular}{|c|c|c|c|c|c|c|} \hline
M & \#CO & $E_\mathrm{b\text{-}n^{th}\,CO/M@Cu(111)}$ & $E_\mathrm{vdW}$ & $z_M$ & magmom \\ \hline
\multirow{5}{*}{Y}
& 0 & -- & --  & 0.792 \AA & 0.000 \\
& 1 & 0.690 eV & 0.293 eV & 0.844 \AA & 0.000 \\
& 2 & 0.712 eV & 0.254 eV & 0.885 \AA & 0.000 \\
& 3 & 0.736 eV & 0.393 eV & 0.925 \AA & 0.000 \\
& 4 & 0.596 eV & 0.500 eV & 1.103 \AA & 0.000 \\ \hline
\multirow{5}{*}{Zr}
& 0 & -- & --  & 0.338 \AA & 0.000 \\
& 1 & 1.192 eV & 0.384 eV & 0.400 \AA & 0.000 \\
& 2 & 1.140 eV & 0.397 eV & 0.411 \AA & 0.000 \\
& 3 & 1.172 eV & 0.559 eV & 0.483 \AA & 0.000 \\
& 4 & 0.708 eV & 0.509 eV & 0.835 \AA & 0.000 \\ \hline
\multirow{5}{*}{Nb}
& 0 & -- & --  & 0.098 \AA & 0.360 \\
& 1 & 1.609 eV & 0.335 eV & 0.078 \AA & 0.000 \\
& 2 & 1.612 eV & 0.599 eV & 0.298 \AA & 0.000 \\
& 3 & 1.529 eV & 0.595 eV & 0.453 \AA & 0.000 \\
& 4 & 0.569 eV & 0.432 eV & 0.915 \AA & 0.000 \\ \hline
\multirow{5}{*}{Mo}
& 0 & -- & --  & 0.110 \AA & 2.200 \\
& 1 & 1.945 eV & 0.378 eV & 0.020 \AA & 0.000 \\
& 2 & 1.769 eV & 0.636 eV & 0.239 \AA & 0.000 \\
& 3 & 1.546 eV & 0.574 eV & 0.519 \AA & 0.000 \\
& 4 & 0.328 eV & 0.314 eV & 1.092 \AA & 0.000 \\ \hline
\multirow{5}{*}{Tc}
& 0 & -- & --  & 0.034 \AA & 1.960 \\
& 1 & 2.190 eV & 0.393 eV & 0.041 \AA & 0.000 \\
& 2 & 1.749 eV & 0.636 eV & 0.301 \AA & 0.000 \\
& 3 & 1.276 eV & 0.547 eV & 0.620 \AA & 0.000 \\
& 4 & 0.051 eV & 0.197 eV & 1.315 \AA & 0.000 \\ \hline
\multirow{5}{*}{Ru}
& 0 & -- & --  & -0.029 \AA & 0.100 \\
& 1 & 2.239 eV & 0.418 eV & 0.100 \AA & 0.000 \\
& 2 & 1.381 eV & 0.642 eV & 0.361 \AA & 0.000 \\
& 3 & 0.876 eV & 0.500 eV & 0.727 \AA & 0.000 \\
& 4 & -0.242 eV & 0.187 eV & 1.434 \AA & 0.000 \\ \hline
\multirow{4}{*}{Rh}
& 0 & -- & --  & -0.001 \AA & 0.000 \\
& 1 & 1.797 eV & 0.414 eV & 0.176 \AA & 0.000 \\
& 2 & 0.825 eV & 0.644 eV & 0.403 \AA & 0.000 \\
& 3 & 0.448 eV & 0.541 eV & 0.711 \AA & 0.000 \\ \hline
\multirow{4}{*}{Pd}
& 0 & -- & --  & 0.089 \AA & 0.000 \\
& 1 & 0.877 eV & 0.392 eV & 0.276 \AA & 0.000 \\
& 2 & 0.471 eV & 0.699 eV & 0.318 \AA & 0.000 \\
& 3 & 0.339 eV & 0.587 eV & 0.450 \AA & 0.000 \\ \hline
\end{tabular}
\caption{Differential binding energy, vdW correction contribution, outward displacement, and magnetic moment (in unpaired electron units) for 4d transition metal substituent atoms (M) on the Cu(111) surface with up to 4 CO molecules adsorbed.}
\label{tab:Cu(111)_4d}
\end{table}